\documentclass[a4paper,11pt]{article}
\pdfoutput=1
\usepackage{amsfonts,latexsym,graphicx,amssymb, amsmath}

\pdfoutput=1

\newcommand{\beq}{\begin{equation}}
\newcommand{\eeq}{\end{equation}}
\newcommand{\beqa}{\begin{eqnarray}}
\newcommand{\eeqa}{\end{eqnarray}}
\newcommand{\beql}{\begin{align}}
\newcommand{\eeql}{\end{align}}

\newcommand{\nn}{\nonumber}

\newcommand{\R}{\mathbb{R}}

\newcommand{\lp}{\left(}
\newcommand{\rp}{\right)}

\newcommand{\bta}{\bar{\tau}}

\DeclareMathOperator\arctanh{arctanh}

\usepackage{jheppub}
\usepackage{graphicx}
\usepackage{subcaption}
\usepackage{dsfont}
\usepackage[dvipsnames]{xcolor}
\usepackage{etoolbox}
\usepackage[T1]{fontenc}
\usepackage{amsmath}
\usepackage{mathtools}
\usepackage{amsthm}
\usepackage{physics}
\makeatletter
\patchcmd{\maketitle}{\@fpheader}{\\}{}{}
\usepackage{stackengine}
\stackMath

\makeatother

\title{Quantum Backreaction on Chronology Horizons}
\author[a,b]{Roberto Emparan}
\author[b]{and Marija Toma\v{s}evi\'c}

\affiliation[a]{Instituci\'o Catalana de Recerca i Estudis Avan\c cats (ICREA),
Passeig Llu\'{\i}s Companys 23, E-08010 Barcelona, Spain}
\affiliation[b]{Departament de F{\'\i}sica Qu\`antica i Astrof\'{\i}sica, Institut de Ci\`encies del Cosmos,
Universitat de Barcelona, Mart\'{\i} i Franqu\`es 1, E-08028 Barcelona, Spain}

\emailAdd{emparan@ub.edu}
\emailAdd{mtomasevic@icc.ub.edu}

\abstract{We extend in two directions our recent investigation of strongly interacting quantum fields in a class of spacetimes with chronology horizons (Misner spacetimes). First, we generalize to arbitrary dimensions the holographic mechanism of chronology protection in the absence of gravitational backreaction. The AdS geometry dual to a conformal field theory in these spacetimes shows, in every dimension, an entire separation between the bulk duals of the chronal and non-chronal regions, with the former being complete, regular geometries. In some instances the protection requires the inclusion of non-planar CFT corrections, which we obtain using double holography. Second, we compute the gravitational backreaction of the quantum fields in the Misner-AdS$_3$ spacetime, and show that the null chronology horizon turns into a strong, spacelike curvature singularity. This is one of the few controlled, explicit examples where we can see quantum effects change a Cauchy horizon into a spacelike singularity.}

\begin{document}

\maketitle

\section{Introduction}
\label{intro}

In a recent article \cite{Emparan:2021xdy} we have used holography to study quantum effects in time machine spacetimes where closed timelike curves are developed. We addressed the question of whether it is possible to send a signal---an excitation of a quantum field---across the chronology horizon, passing into  the non-chronal region where CTCs are present. Quantum field theory in these situations has been studied since long ago (see e.g., \cite{Hiscock:1982vq,Morris:1988tu, PhysRevD.43.3878, PhysRevD.46.603, Klinkhammer:1992tb, Boulware:1992pm, PhysRevLett.71.2517, Visser:1995cc,Kay_1997,Li_1998,Li_1999,Alonso-Serrano:2021ujk}), typically solving for free fields in simple spacetimes with chronology horizons. It is found that the quantum stress-energy tensor in the ground state of the field develops a divergence as the chronology horizon is approached, which leads to the speculation that, by coupling the field to gravity, a strong backreaction occurs that turns the chronology horizon into a curvature singularity. At this point,  quantum gravity will rule the crossing, hopefully preventing it.

Our results in \cite{Emparan:2021xdy} indicate that simpler, more controllable physics may protect the normal chronological order. We employed the AdS$_3$/CFT$_2$ correspondence to study strongly coupled quantum fields in a suitable class of two-dimensional time machines, known as Misner spacetimes \cite{Misner:1965zz}. The idea is that, for the stress tensor divergence to prevent a signal---such as a wave packet of the quantum field---from crossing the chronology horizon, there must be some interaction between the signal and the background field state. Of course, ultimately the gravitational interaction is present and couples all the fields to each other. But the dynamics that is then induced often runs into hard-to-solve, even uncontrollable (e.g., Planck-scale) physics. So we took a minimalist approach where the quantum fields interact with themselves but not with gravity, and examined the consequences.

We found that the holographic CFT reproduced the divergence of the stress-energy tensor at the chronology horizon. But, more interestingly, we also found that the self-interaction of the fields, without any gravitational backreaction, manages to banish the passage of quantum excitations of the field into the non-chronal region. In this instance, quantum gravity need not be invoked to avoid abnormal chronology.

The protection mechanism we found has an appealing dual description. The three-dimensional bulk geometry that describes the two-dimensional CFT splits into two disconnected bulk spacetimes, without and with CTCs, with the two regions connected only on the asymptotic boundary. Then, an excitation of the conformal fields will dually propagate in the bulk without ever encountering any singularity, and not arriving at the chronology horizon in any finite proper time. An observer in the CFT sees the excitation interact with the field state, increasing its own energy without bound as it approaches the chronology horizon, and never crossing it.  Thus the non-chronal region, with CTCs, is physically split from the normal chronal region, even though the background Misner spacetime remains unaffected. The bulk dual of the chronal region is smooth and complete, but in the non-chronal region it is singular, so the CFT is ill-defined there. 

In this article we extend the analysis of \cite{Emparan:2021xdy} in two directions. First, we show that the mechanism of chronology protection that we found is not an artifact of low-dimensional physics. We build holographic duals of strongly coupled CFTs in higher-dimensional generalizations of the Misner spacetime, and find that the same complete split occurs between the bulk duals of the chronal and non-chronal regions. The bulk in the chronal region is complete and regular in every dimension, so the absence of curvature singularities in the bulks of \cite{Emparan:2021xdy} was not a consequence of having only three dimensions. Furthermore, we will prove an assertion that we made in \cite{Emparan:2021xdy}, namely, that non-planar corrections of the CFT suffice to eliminate potential violations of chronology between two entangled time machines. In this case, we find that the chronal bulk is not geodesically complete but terminates at a future singularity.  We prove this through the use of `double holography'. 

Our second result is perhaps farther-reaching. One may worry that `dark' systems that propagate in the Misner spacetime, but do not interact with the CFT---so they can be regarded as not propagating in the bulk spacetime, but only on the boundary geometry---would find no problem crossing the chronology horizon. But, as we mentioned, the universality of gravitational coupling should extend the protection to every other system. In other words, we are back to the conjecture that the gravitational backreaction of the quantum fields should close up the chronology horizon passage.

Previous attempts at this problem faced the difficulty that the strong quantum backreaction that is sought is incompatible with the perturbative methods often used, since these assume that the effects are small. To overcome this obstacle, we resort to a `braneworld holography' approach that allows to deal with the finite gravitational backreaction of the CFT \cite{Verlinde:1999fy,Gubser:1999vj,Karch:2000ct,Emparan:2002px}. We will then show that the chronology horizon is replaced by a strong curvature singularity, which is spacelike---implementing strong cosmic censorship---and severs apart, once and for all, the chronal and non-chronal regions. This is among the few controlled, explicit examples of a Cauchy horizon being turned into a spacelike singularity through quantum effects.

In the next section we introduce the class of time machine Misner-AdS spacetimes in arbitrary dimension that we consider. In Sec.~\ref{sec:holobulks} we construct the  bulk duals for a strongly coupled conformal field theory propagating in those backgrounds. In Sec.~\ref{sec:quMisner} we obtain the gravitational backreaction of these fields on the geometry of Misner-AdS$_3$ and study the structure if the singularity that appears. In Sec.~\ref{sec:doubleh} we use double holography to compute quantum corrections to the bulk dual of the CFT in Misner-AdS$_2$, and thus prove a claim made in \cite{Emparan:2021xdy} about an extension of chronology protection. Sec.~\ref{sec:discus} summarizes and discusses the outlook of our results.

\section{Misner-AdS$_{n+2}$ time machines}\label{sec:misnerads}

We refer to Misner-AdS spacetimes as the class of geometries in $n+2$ dimensions ($n\geq 0$) described by the metrics
\begin{equation}\label{misneradsn2}
   ds^2 = -(dt+ t d\psi)^2 +\sum_{i=1}^n (dy_i+ y_i d\psi)^2+d\psi^2\,,
\end{equation}
with $t$ and $y^i$ running along all of $\R$, and $\psi$ periodically identified at fixed $t$ and $\mathbf{y}=(y^1,\dots,y^n)$ as
\beq
(t,\mathbf{y},\psi)\sim (t,\mathbf{y},\psi +\Delta)\,.
\eeq
Since
\begin{equation}
    g_{\psi\psi}=1-t^2+\mathbf{y}^2\,,
\end{equation}
we see that the $\psi$ coordinate becomes timelike when  $t^2>1+\mathbf{y}^2$. Then the surface $t^2-\mathbf{y}^2=1$ is a chronology horizon,\footnote{There are two horizons, $t=\pm \sqrt{1+\mathbf{y}^2}$, but we are mostly interested in the future one.} and closed timelike curves appear when it is crossed from the chronal region to the non-chronal region beyond. The coordinates $(t,\psi,\mathbf{y})$ extend analytically the manifold across this surface, but there are incomplete null geodesics that terminate at it, while others cross it smoothly (see \cite{Emparan:2021xdy} for explicit details). One then refers to it as a quasi-regular surface. Although sometimes spacetime is taken to terminate at this surface, we are interested in keeping the extension into the non-chronal, time-machine region.

The case $n=0$, where the horizon is at $t^2=1$,  was thoroughly studied in \cite{Emparan:2021xdy}. When $n\geq 1$ the chronology horizon is non-compactly generated. The Misner-AdS$_4$ spacetimes were introduced in \cite{Li_1999}.

\subsection*{Global structure}

It is easy to verify that the curvature of \eqref{misneradsn2} is everywhere equal to $-1$. Therefore, the spacetime must be locally equivalent to unit-radius AdS$_{n+2}$, and allow a description with AdS$_{n+2}$ as its covering space. To see this, begin with the representation of AdS as the hyperboloid
\beq\label{hyperads}
-V^+V^- -W^2+\mathbf{Y}^2=-1\quad\text{in}\quad ds^2=-dV^+ dV^--dW^2+d\mathbf{Y}^2\,,
\eeq
where $\mathbf{Y}=(Y^1,\dots,Y^n)$.
The metric \eqref{misneradsn2} is obtained by setting 
\begin{align}
V^+&=e^\psi\,,\\
V^-&=e^{-\psi}\lp 1-t^2+\mathbf{y}^2\rp\,,\\ 
W&=t\,,\\
\mathbf{Y}&=\mathbf{y}\,.
\end{align}
Therefore, Misner-AdS$_{n+2}$ can be defined as the hyperboloid \eqref{hyperads} with points identified under
\begin{equation}\label{idV+}
V^+\sim e^\Delta V^+\,,\qquad V^-\sim e^{-\Delta}V^-\,,
\end{equation}
at fixed $W$ and $Y^i$, and restricted to the half-space
\begin{equation}\label{V+pos}
    V^+>0\,.
\end{equation}

The identifications \eqref{idV+} are made on surfaces where $V^+ V^-$ is constant. Then, given the restriction to \eqref{V+pos}, the related points are spacelike-separated where $V^->0$ (chronal region), and timelike-separated, creating CTCs, where $V^-<0$ (non-chronal region). The surface $V^-=0$ is a chronology horizon. 

It will be useful to employ other coordinates where the metric becomes diagonal. First, we make
\begin{equation}
   W =\tau \cosh\chi\,,\qquad Y^i= \tau \sinh\chi\,\omega^i\,,
\end{equation}
where $\omega^i$ are direction cosines of $S^{n-1}$, such that $\sum_i\omega^i\omega^i=1$.
In the chronal region, where $\tau^2<1$,  we set
\begin{equation}
    V^\pm =\sqrt{1-\tau^2}\,e^{\pm\phi}\,,
\end{equation}
and in the non-chronal region, where $\tau^2>1$,
\begin{equation}
    V^\pm =\pm \sqrt{\tau^2-1}\,e^{\pm\phi}\,.
\end{equation}
The coordinate $\phi$ has periodicity $\Delta$.
Although the resulting metric is not defined on the chronology horizon, elsewhere it takes the form
\begin{equation}\label{misneradsnpatch}
    ds^2=-\frac{d\tau^2}{1-\tau^2}+(1-\tau^2)d\phi^2+\tau^2 \lp d\chi^2+\sinh^2\chi\, d\Omega_{n-1}\rp\,.
\end{equation}
The last factor is the hyperbolic space $H_{n}$. This will be the representation of choice for Misner-AdS$_{n+2}$ for most of this article.

To study the geometry near the (future) chronology horizon, we change
\begin{equation}
    \tau=1+\epsilon\, T\,,\qquad \chi=\sqrt{\epsilon}\,\xi
\end{equation}
and expand to leading order in $\epsilon\ll 1$, to find
\begin{equation}
    \epsilon^{-1} ds^2=\frac{dT^2}{2T}-2Td\phi^2+d\xi^2+\xi^2 d\Omega_{n-1}\,.
\end{equation}
This is the direct product of the two-dimensional flat Misner spacetime \cite{Misner:1965zz}, with coordinates $(T,\phi)$ and CTCs for $T>0$, and the plane $\R^n$. For the same reasons that Rindler spacetime is the universal geometry near a black hole horizon, the local structure of these chronology horizons always reduces to Misner space on short scales.

To get to this limit in the coordinates of \eqref{misneradsn2}, we take
\begin{equation}
    t=1+\epsilon\, \bar{T}\,,\qquad y^i=\sqrt{\epsilon}\,\xi\, \omega^i
\end{equation}
and send $\epsilon\to 0$ to find
\begin{equation}
    \epsilon^{-1} ds^2=-2d\bar{T}\,d\psi-2\bar{T}d\psi^2+\lp d\xi+\xi d\psi\rp^2+\xi^2 d\Omega_{n-1}\,.
\end{equation}
It may not be apparent, but this is again a flat space metric, now with CTCs where $\bar{T}>\xi^2/2$.

Finally, although we will not make use of it, it is interesting to obtain another form of Misner-AdS$_{n+2}$ in terms of Poincar\'e coordinates.  Setting
\begin{equation}
    z=\frac1{V^+}\,,\qquad \eta=-\frac{W}{V^+}\,,\qquad \mathbf{x}=\frac{\mathbf{Y}}{V^+}\,,
\end{equation}
we have that Misner-AdS$_{n+2}$ corresponds to
\begin{equation}
    ds^2=\frac{dz^2-d\eta^2+d\mathbf{x}^2}{z^2}
\end{equation}
with the identifications
\begin{equation}
    (z,\eta,\mathbf{x})\sim e^\Delta (z,\eta,\mathbf{x})\,.
\end{equation}
These make CTCs appear whenever $\eta^2>z^2+\mathbf{x}^2$.

\subsection*{Misner-AdS$_3$  $\supset$ static BTZ}

In the three-dimensional case, $n=1$, the identifications along AdS boosts \eqref{idV+} are the same as for the static BTZ black hole \cite{Banados:1992wn,Banados:1992gq}. The two solutions differ only in the ranges they cover: Misner-AdS$_3$ includes all of $V^-\in\R$, covering both sides of the chronology horizon $V^-=0$, but for BTZ one customarily keeps only the chronal region $V^->0$. This is often implemented by choosing adapted coordinates $(\chi,r,\phi)$, which are defined, outside the horizon where $r^2>1$, by
\begin{equation}
    V^\pm =r e^{\pm\phi}\,,\qquad W^\pm =\pm\sqrt{r^2-1} e^{\pm \chi}
\end{equation}
(with $W^\pm=W\pm Y)$ and, inside the horizon where $r^2<1$, by
\begin{equation}
     V^\pm =r e^{\pm\phi}\,,\qquad W^\pm =\sqrt{1-r^2} e^{\pm \chi}\,.
\end{equation}
Then, only the region where $V^\pm\geq 0$ is kept. The metric away from the horizon is
\begin{equation}\label{BTZ}
    ds^2=-(r^2-1)d\chi^2+\frac{dr^2}{r^2-1}+r^2 d\phi^2\,,
\end{equation}
with $\chi$ the asymptotic time coordinate, and with the periodicity of $\phi$ determining the black hole mass.

In hyperboloid coordinates, the future and past black hole horizons are at $W^+=0$ and $W^-=0$.
These horizons are also present in the Misner-AdS$_3$ spacetime \eqref{misneradsn2}, but they are not manifest. They are the null surfaces at $t\pm y=0$. So in Misner-AdS$_3$ the chronology horizon is behind a black hole horizon that causally separates it from the asymptotic observers in the chronal region. This does not happen in Misner-AdS$_2$.

The Misner chronology horizon is the locus $r=0$ in the BTZ solution. As we indicated above, rather than viewing it as a singularity where spacetime ends, we take it as a quasi-regular null surface that one can cross into the non-chronal region.\footnote{In the BTZ metric \eqref{BTZ}, the $(r,\phi)$ coordinates become singular at $r=0$ but in an extendible way since the geometry there, $\sim -dr^2+r^2 d\phi^2$, is flat. By contrast, in the Schwarzschild black hole near $r=0$ one gets $\sim -rdr^2+r^2 d\phi^2$, which is part of a Kasner singularity (see Sec.~\ref{sec:curvsing}).} Starting from the BTZ representation \eqref{BTZ}, this is conveniently attained by changing $r^2=1-\tilde{r}^2$, so the metric now takes the form
\begin{equation}\label{Rads}
    ds^2=-(\tilde{r}^2-1) d\phi^2+\frac{d\tilde{r}^2}{\tilde{r}^2-1} +\tilde{r}^2d\chi^2\,.
\end{equation}
The region $\tilde{r}>1$ is the Rindler-AdS$_3$ spacetime, with $\chi$ running along $(-\infty,\infty)$ and with periodic time $\phi$. The acceleration horizon $\tilde{r}=1$ is the chronology horizon.

We conclude that Misner-AdS$_3$ consists of a chronal BTZ black hole connected through a chronology horizon to a non-chronal Rindler-AdS$_3$ with periodic time.

\subsection*{Misner-AdS$_{n+2}$ $\supset$ constant curvature black holes}

The previous analysis can be generalized to show that, for all $n\geq 1$, Misner-AdS$_{n+2}$ has a black hole horizon in addition to the chronology horizon. Refs.~\cite{Aminneborg:1996iz,Banados:1997df} observed that the discrete quotient of AdS$_{n+2}$ by boost orbits yields a black hole geometry that generalizes the BTZ solution. That is,  Misner-AdS$_{n+2}$ has these constant curvature black holes in the chronal region, while the non-chronal region is again a periodic-time Rindler-AdS$_{n+2}$. As black hole solutions, these geometries present a number of peculiarities \cite{Aminneborg:1996iz,Banados:1997df,Banados:1998dc,Aminneborg:2008sa} which we will not discuss, since it is their interpretation as time machine backgrounds where quantum fields propagate that matters most to us.

\section{Holographic bulk dual of Misner-AdS$_{n+2}$}\label{sec:holobulks}

We now proceed to construct AdS bulk geometries in $n+3$ dimensions that have the geometry \eqref{misneradsnpatch} of Misner-AdS$_{n+2}$ at the conformal boundary. 
\subsection{Bulk construction}

Our strategy is an extension of the one used in \cite{Hubeny:2009rc} to find a four-dimensional AdS bulk with the BTZ black hole at its boundary. Following our previous discussion, that spacetime can be equivalently seen as a bulk with the chronal region of Misner-AdS$_3$ at the boundary. We have found how to generalize the construction to higher dimensions, and also extended it to yield the bulk for the non-chronal Rindler-AdS region.

It will be convenient to introduce a parameter
\begin{equation}\label{kappasign}
    \kappa=\rm{sign}(\tau^2-1)=
    \begin{cases}
    +1\quad &\text{non-chronal region\,,}\\
    -1\quad &\text{chronal region\,.}
    \end{cases}
\end{equation}

\paragraph*{Chronal bulk.} Let us first obtain a spacetime whose boundary is the geometry \eqref{misneradsnpatch} with $\tau^2<1$.

Consider the solution for the spherical black hole in AdS$_{n+3}$, with the metric on $S^{n+1}$ written in a particular way, namely, we represent the polar angle by $\arccos(1/r)$, so that the solution takes the form
\begin{equation}\label{schwads}
    ds^2=-F_-(\rho)dT^2+\frac{d\rho^2}{F_-(\rho)}+\rho^2\lp\frac{dr^2}{r^2(r^2-1)}+\frac{r^2-1}{r^2}(d\theta^2+\sin^2\theta d\Omega_{n-1})\rp \,,
\end{equation}
with $1\leq r\leq \infty$. The function $F_-(\rho)$ is given by
\begin{equation}\label{Fkappa}
    F_\kappa(\rho)=\rho^2-\kappa+\kappa\frac{\alpha}{\rho^n}
\end{equation}
with $\kappa=-1$.

Now we double Wick rotate
\begin{equation}
    T\to i\phi\,,\qquad \theta\to i\chi\,,
\end{equation}
to find
\begin{equation}\label{wickschwads}
    ds^2=\frac{\rho^2}{r^2}\lp-\frac{dr^2}{1-r^2}+(1-r^2)(d\chi^2+\sinh^2\chi d\Omega_{n-1})\rp +F_-(\rho)d\phi^2+\frac{d\rho^2}{F_-(\rho)}\,.
\end{equation}
We now take $r^2\leq 1$ in order to have the right signature, and change to
\begin{equation}
    r^2=1-\tau^2\,,
\end{equation}
with $0\leq \tau^2<1$.
The resulting metric can now be written as 
\begin{equation}\label{lowerbulk}
    ds^2 = \frac{\rho^2}{1-\tau^2} \bigg( - \frac{d\tau^2}{1-\tau^2}+\tau^2 \lp d\chi^2+\sinh^2\chi\, d\Omega_{n-1}\rp  + (1-\tau^2)\frac{F_-(\rho)}{\rho^2} d\phi^2\bigg) + \frac{d\rho^2}{F_-(\rho)}.
\end{equation}
At the boundary $\rho\to\infty$, taking the conformal factor $\rho^2/(1-\tau^2)$, we recover the Misner-AdS$_{n+2}$ geometry \eqref{misneradsnpatch}, as we desired.

If $\alpha<0$ then when $n\geq 1$ there is a naked curvature singularity at $\rho=0$, which indictaes a pathological state.\footnote{Instead, for $n=0$ and $\alpha<0$ the geometry \eqref{lowerbulk} is a BTZ black hole with horizon at $\rho=0$. These were identified in \cite{Emparan:2021xdy} as entangled CFT states.} Thus we will require that $\alpha\geq 0$, and furthermore that its value is fixed by the periodicity $\Delta$ of $\phi$, in such a way that there are no conical singularities at $\rho=\rho_0$, the largest positive root of $F_-$. Setting
\begin{equation}\label{alpharho-}
    \alpha=\rho_0^n(\rho_0^2+1)\,,
\end{equation}
the condition is
\begin{equation}\label{Delrho-}
    \Delta=\frac{4\pi}{F_-'(\rho_0)}=\frac{4\pi\rho_0}{(n+2)\rho_0^2+n}\,.
\end{equation}
When this is satisfied, we obtain a complete, non-singular bulk solution. Then, the CFT state in Misner-AdS is regular and well defined for all $\tau^2<1$.

It is peculiar that, for all $n\geq 1$, there is a maximum value of $\Delta$,
\begin{equation}
    \Delta_{max}= \frac{2\pi}{\sqrt{n(n+2)}}
\end{equation}
beyond which there is no bulk dual of this type.\footnote{This, and other observations that follow, are the double Wick rotation of the lower bound on the temperature of spherical AdS black holes, and of the coexistence of large and small AdS black hole phases.}  For larger $\Delta$, the only known regular bulk solution corresponds to $\alpha=0$. The latter is the geometry of AdS$_{n+3}$, but since we require that $\phi$ is periodic, it is actually a constant curvature black hole with horizon at $\rho=0$. Since it has two asymptotic regions, and hence two entangled CFTs, it is unclear whether we should consider it alongside states of a single CFT. In any case, we interpret that for $\Delta>\Delta_{max}$ the rate of evolution towards the time machine seems to be so fast, compared to the energy gap in the $\phi$ circle, that the CFT cannot keep up and remains unexcited.

For $\Delta<\Delta_{max}$ there are two bulks with the same $\Delta$ but different non-zero values of $\alpha$. The relevance of Euclidean calculations in time-dependent configurations such as these may be questioned, but the actions of the Euclideanized black holes \eqref{schwads} suggest that, for a given $\Delta<\Delta_{max}$, the state with the larger $\alpha$ should dominate.\footnote{If it makes sense to compare single-boundary and two-boundary configurations through their Euclidean actions, then the bulks \eqref{lowerbulk} with the largest $\alpha$ are presumably preferred when $\Delta<\Delta_{HP}=\frac{2\pi}{n+1}$ (with $\Delta_{HP}$ the `Hawking-Page periodicity'). For $\Delta>\Delta_{HP}$ the preferred bulk would have $\alpha=0$.  These features were absent for the case $n=0$ studied in \cite{Emparan:2021xdy}.}

Let us emphasize that these are not thermal states since there is no horizon in the bulk.  Instead, the CFT is in a confined phase, and remains so throughout the entire time evolution, up until the future chronology horizon at $\tau\to 1^-$. In the bulk, the latter is reached at infinite proper time.

\paragraph*{Non-chronal bulk.} In order to find a spacetime whose boundary  is \eqref{misneradsnpatch} with $\tau^2>1$, we take the metric of the hyperbolic black hole in AdS$_{n+3}$, written in the form
\begin{equation}\label{hyperbhads}
    ds^2= -F_+(\rho)d\phi^2+\frac{d\rho^2}{F_+(\rho)}+\rho^2\lp\frac{dr^2}{r^2(r^2+1)}+\frac{r^2+1}{r^2}(d\chi^2+\sinh^2 \chi\, d\Omega_{n-1})\rp\,,
\end{equation}
with $F_+(\rho)$ given by \eqref{Fkappa} with $\kappa=+1$, and $0< r < \infty$. The `orbital space' within the big brackets is the hyperbolic space $H_{n+1}$ (the proper radial distance is $\mathrm{arcsinh}(1/r)$). For continuity with \cite{Emparan:2021xdy}, we have chosen $\alpha<0$ to correspond to positive mass black holes. We are \emph{not} performing any Wick rotation, but we have renamed the time coordinate as $\phi$, since we are identifying it periodically and thus introducing CTCs. 
Now making
\begin{equation}
    r^2=\tau^2-1
\end{equation}
with $\tau^2> 1$, we rewrite the metric as
\begin{equation}\label{upperbulk}
    ds^2 = \frac{\rho^2}{\tau^2-1} \bigg( \frac{d\tau^2}{\tau^2-1}+\tau^2 \lp d\chi^2+\sinh^2 \chi\, d\Omega_{n-1}\rp  - (\tau^2-1)\frac{F_+(\rho)}{\rho^2} d\phi^2\bigg) + \frac{d\rho^2}{F_+(\rho)}\,.
\end{equation}
At the asymptotic boundary $\rho\to\infty$, with the conformal factor $\rho^2/(\tau^2-1)$, this yields Misner-AdS$_{n+2}$ with $\tau^2>1$, as we desired.

Real positive zeroes of $F_+$, when they exist, are horizons of the hyperbolic black hole, and since $\phi$ is periodic, they correspond to chronology horizons in the bulk. 

When $\alpha<0$ there is always a positive root of $F_+$ and the solution is a positive-mass hyperbolic black hole. 
If $\alpha>0$ the mass of the hyperbolic black hole is negative, but it is known that when $n\geq 1$ there exists a range of negative masses, $0<\alpha\leq\alpha_{ext}$, with
\begin{equation}\label{alphaext}
    \alpha_{ext}=\frac{2}{n+2}\left(\frac{n}{n+2}\right)^{n/2}
\end{equation}
for which the spacetime is a black hole with a regular horizon \cite{Emparan:1999gf}. 

Thus, in the range $\alpha\leq \alpha_{ext}$ the bulk is a hyperbolic black hole with periodic time, and as we explained in \cite{Emparan:2021xdy} and will discuss below, the state of the CFT has a thermal component, due to the entanglement between two CFTs in two asymptotic boundaries. The black hole exterior is non-chronal, but the interior contains a chronal region. At the bound \eqref{alphaext}, the horizon is extremal, and both the exterior and the interior are non-chronal.  

When $\alpha\neq 0$ (and $n\geq 1$) the interior ends at a curvature singularity at $\rho=0$. This is different than what was found in \cite{Emparan:2021xdy} for $n=0$, where one could cross $\rho=0$ in the interior to emerge in a BTZ exterior. Thus when $n\geq 1$ the wormhole connecting the two asymptotic boundaries is not traversable. The only case where the interior does not close off at a curvature singularity is the state with $\alpha=0$, where the chronal bulk is a constant curvature black hole connected across $\rho=0$ to the non-chronal exterior of a Rindler-AdS$_{n+2}$ with periodic time. 

Finally, the bulks where $\alpha>\alpha_{ext}$ are pathological since the curvature singularity at $\rho=0$ is naked. 

There is in general no relation between $\alpha$ and $\Delta$, at least in the Lorentzian geometry. However, in the Euclidean section with imaginary $\phi$, if we require that the Lorentzian periodicity $\Delta$ be the same as the Euclidean (thermal) periodicity $\beta=2\pi/T_H$ of the black hole outer horizon, then
\begin{equation}\label{Delrho+}
    \Delta=\frac{4\pi}{|F_+'(\rho_0)|}=\frac{4\pi\rho_0}{|(n+2)\rho_0^2-n|}\,,
\end{equation}
where $\rho_0$ is the largest positive root of
\begin{equation}\label{alpharho+}
    \alpha=-\rho_0^n (\rho_0^2-1)\,,
\end{equation}
with $\alpha\leq \alpha_{ext}$. When $\alpha=0$ at $\rho_0=1$, we have $\Delta=2\pi$, while when $\alpha=\alpha_{ext}$ at $\rho_0^2=n/(n+2)$ the periodicity diverges, $\Delta\to\infty$.

\subsection{Holographic stress tensor} 

Putting together the results of the previous constructions, the bulks in both regions, $\kappa=\pm 1$, can be written in a unified form as 
\begin{equation}\label{kappabulk}
    ds^2 = \frac{\rho^2}{|\tau^2-1|} \bigg( \frac{d\tau^2}{\tau^2-1}+\tau^2 \lp d\chi^2+\sinh^2 \chi\, d\Omega_{n-1}\rp  - (\tau^2-1)\frac{F_\kappa(\rho)}{\rho^2} d\phi^2\bigg) + \frac{d\rho^2}{F_\kappa(\rho)}\,,
\end{equation}
with $F_{\kappa}(\rho)$ in \eqref{Fkappa}.
This is valid for all $n\geq 0$. The case $n=0$, where the hyperbolic space $H_n$ is absent and $F(\rho)$ is finite at $\rho=0$, was discussed at length in \cite{Emparan:2021xdy}. There, the method of AdS/CFT bulk reconstruction was used, which yielded the same geometries as here but in the Fefferman-Graham gauge.

The holographic CFT stress-energy tensor is most easily obtained from the one for the AdS black holes \cite{Emparan:1999gf}, changing coordinates as above, and applying a Weyl transformation for the conformal factor at the boundary. When $n$ is even it contains an anomalous trace, which we are not interested in, and furthermore there are traceless contributions that are independent of $\alpha$, which we also subtract. This is then equivalent to performing background subtraction with the reference state $\alpha=0$, which is locally empty AdS. The result in $(\tau,\phi,\chi,\Omega)$ coordinates is
\begin{align}\label{holostress}
    \langle T_a{}^b \rangle&= \frac{1}{8 \pi G} \frac{-\kappa\alpha}{|1 - \tau^2|^{\frac{n+2}{2}}}\, \text{diag}(1,-(n+1),1,\dots,1)\nn\\
    &= \frac{1}{8 \pi G} \frac{\alpha}{(1-\tau^2)|1 - \tau^2|^{\frac{n}{2}}}\, \text{diag}(1,-(n+1),1,\dots,1)
\end{align}
The most salient feature is the divergence at the chronology horizon. In four dimensions $(n=2)$,  \cite{Li_1999} found the same divergence for a free conformal scalar field. In two dimensions $(n=0)$, \eqref{holostress} gives the same result as in \cite{Emparan:2021xdy}.

States with $\alpha>0$ in the chronal region $\kappa=-1$ have negative energy density and positive tension along $\phi$, 
\begin{equation}
    -\langle T_\tau^\tau \rangle<0\,,\qquad -\langle T_\phi^\phi \rangle>0\,.
\end{equation}
This was explained in \cite{Emparan:2021xdy} as originating from the Casimir effect in the $\phi$ circle. Indeed, near $\tau=0$ the stress tensor is that of the CFT in a static cylinder \cite{Myers:1999psa}.

In the non-chronal region, the states with $\alpha<\alpha_{ext}$ contain a mixture of Casimir and thermal energies, but only when $\alpha<0$ the thermal component (of thermofield-double type) dominates, and makes the energy density $-\langle T_\phi^\phi\rangle$ and all the pressures positive. 

\subsection{Holographic chronology protection}\label{sec:holchro}

Let us then summarize our results for the bulk geometries when $n\geq 1$: 
\begin{enumerate}
    \item \textit{Chronal region}. When $\alpha>0$ the bulk is regular and complete, corresponding to sensible CFT states (besides the divergence at the chronology horizon) with negative Casimir energy. When $\alpha<0$ the bulk has naked curvature singularities, hence the CFT is ill-defined everywhere. 
    \item \textit{Non-chronal region}. When $\alpha<\alpha_{ext}$ the bulks are hyperbolic black holes with periodic time. Therefore, despite the presence of CTCs, the entanglement with a second CFT apparently allows the quantum fields to be well defined (again, away from the chronology horizon). For $\alpha>\alpha_{ext}$ the bulks have naked curvature singularities, so the CFT is ill-defined.
\end{enumerate}

Point 1 implies that we recover the main conclusion of \cite{Emparan:2021xdy}, namely, chronology protection by splitting the non-chronal bulk region and the chronal one as separate spacetimes. It is only on the boundary that one can cross the chronology horizon at $\tau^2=1$. The chronal bulk geometry is geodesically complete, and the horizon $\tau=1^-$ lies at its asymptotic future boundary.

Suppose, then, that we start at $\tau=0$ in a regular state of the CFT in Misner-AdS$_{n+2}$, hence a bulk with $\alpha>0$. Any excitation of the CFT on top of this state dually propagates in the bulk geometry, extending to arbitrarily large values of its proper time without ever crossing into a non-chronal region. In the CFT view, as the excitation approaches the chronology horizon of Misner-AdS$_{n+2}$, its energy grows unbounded owing to its interaction with the background field state, which diverges there, and it never manages to cross to the non-chronal region.

Point 2 indicates the possibility of sensible states of the CFT in a non-chronal spacetime. These are entangled states of two CFTs, which lie on separate Misner-AdS$_{n+2}$ spacetimes at the asymptotic boundaries of the hyperbolic black hole. Whenever $\alpha\neq 0$ there is a singularity in the interior that makes this system a non-traversable wormhole.

When $\alpha=0$ we can write the solution as AdS$_{n+3}$ foliated by Misner-AdS$_{n+2}$ slices, i.e., 
\begin{equation}\label{zeroalpha}
    ds^2=d\sigma^2+\cosh^2\sigma\, \left[ds^2(\text{Misner-AdS$_{n+2}$})\right]\,.
\end{equation}
As we already noted, the identifications in Misner-AdS$_{n+2}$ make \eqref{zeroalpha} a constant curvature black hole. Allowing the extension across its chronology horizon, it is possible to pass from the chronal region in one boundary, say $\sigma\to\infty$, to the non-chronal region in the other boundary,  $\sigma\to -\infty$. The wormhole is then traversable.\footnote{When $n\geq 1$ only the state $\alpha=0$ has this interpretation, but when $n=0$ the states with $\alpha<1/2$ (including negative values) are also traversable wormholes \cite{Emparan:2021xdy}.}  In \cite{Emparan:2021xdy} we explained that such violations of chronology should be artifacts of the leading large-$N$ limit of the CFT, which will be banished when $1/N^2$ corrections are included.  In Sec.~\ref{sec:doubleh} we will manage to prove this point when $n=0$. 

From the boundary viewpoint, it is more sensible to describe the CFT states for a given value of $\Delta$ in the Misner-AdS geometry. These are:
\begin{enumerate}
    \item \textit{Chronal region}. For $\Delta>\Delta_{max}$ there is no single-CFT state with non-zero stress tensor. We interpret that when $\Delta$ is very large, the rate of acceleration towards the time machine (relative to the length of the $\phi$ circle) is too fast for the CFT to respond. The state with zero stress tensor $(\alpha=0)$ seems to exist only as a thermofield-double type of state, where the thermal energy exactly cancels the Casimir energy. Its Lorentzian section can have arbitrary $\Delta$, but the Euclidean one must have $\Delta=2\pi$.
    
    When $\Delta<\Delta_{max}$ there exist two states of a single CFT with non-zero stress, with different values of $\alpha$ given by \eqref{alpharho-}, \eqref{Delrho-}. The state with the largest $\alpha$ possibly dominates.
    
    \item \textit{Non-chronal region}. If we require regularity in the Euclidean section with the same periodicity for imaginary and real time, then for every value of $\Delta$ there exist two states, with different $\alpha<\alpha_{ext}$, that satisfy \eqref{Delrho+} and \eqref{alpharho+}. For $\Delta<2\pi$ one of the states has $\alpha>0$ and the other $\alpha<0$, corresponding, respectively, to hyperbolic black holes of negative and positive mass. For $\Delta>2\pi$ both of them have $\alpha>0$ (i.e. negative mass). The state with the smallest $\alpha$ (largest bulk black hole) is expected to be dominant.
\end{enumerate}

\section{Quantum backreaction in Misner-AdS$_3$}\label{sec:quMisner}

In our previous study, the CFT lived on the fixed background geometry that lies at the non-dynamical boundary of an AdS$_{n+3}$ spacetime. In order to study the backreaction of the quantum CFT  we have to make the AdS boundary dynamical. For this purpose, we will employ the framework of braneworld holography \cite{Verlinde:1999fy,Gubser:1999vj,Karch:2000ct,Emparan:2002px}, in the particular case of $n=1$ where we can find exact solutions for the four-dimensional bulk. In these setups, a brane is introduced in an asymptotically AdS bulk spacetime at a finite distance from the boundary, thus cutting off part of the asymptotic region. According to the holographic dictionary, the metric induced on the brane, $h_{ab}$, in three dimensions and with a negative cosmological constant, solves the semiclassical equations
\beq\label{3Deffeqn}
R_{ab}-\frac12 h_{ab} \lp R+2\rp+\dots =8\pi G_ 3\langle T_{ab} \rangle\,,
\eeq
where the dots denote higher curvature terms that are suppressed by even powers of the parameter $\ell$ (introduced below) that measures the position of the brane (in Fefferman-Graham gauge), and which also characterizes the strength of the gravitational backreaction of the CFT. These terms are generated by the integration of ultraviolet degrees of freedom of the CFT above the energy cutoff $\sim 1/\ell$. Having a bulk solution that is exact in $\ell$ resums all these corrections. The renormalized stress tensor in the right-hand side of \eqref{3Deffeqn} is that of the holographic CFT below the cutoff.

Therefore, braneworld holography allows to solve the problem of the quantum backreaction of a conformal theory on the semiclassical geometry where it propagates, beyond the regime of small perturbations. For more details, applied to the solutions that we use, we refer the reader to \cite{Emparan:2020znc}.

\subsection{Bulk dual}

Refs.~\cite{Emparan:2002px,Emparan:2020znc} described a braneworld solution of the Einstein-AdS equations in four dimensions (following \cite{Emparan:1999wa,Emparan:1999fd}) that gives the  bulk dual of the BTZ black hole with quantum backreaction. Given our identification of BTZ as part of the Misner-AdS$_3$ spacetime, we can use the same bulk solution to investigate the backreaction of quantum fields on the Misner chronology horizon.

The bulk metrics we need are the AdS C-metrics\footnote{Relative to the conventions in \cite{Emparan:2020znc}, for consistency with the previous sections we have renamed $\mu\to - \kappa\alpha$, $t\to \chi$,  $x\to 1/\rho$, and considered both values of $\kappa$. The cosmological radius on the brane is $\ell_3=1$. For small $\ell$ we have $\ell_4\simeq \ell$.} 
\beq\label{statc2}
ds^2=\frac{\ell^2}{\lp \ell\rho+r\rp^2}\lp\rho^2\lp \kappa H_\kappa(r) d\chi^2+\frac{dr^2}{H_\kappa(r)}\rp
+r^2\lp \frac{d\rho^2}{F_\kappa(\rho)}-\kappa F_\kappa(\rho)d\phi^2\rp\rp\,,
\eeq
where $F_\kappa(\rho)$ is the same as in \eqref{Fkappa} with $n=1$, that is
\beq\label{FHkappa1}
F_\kappa(\rho)= \rho^2-\kappa +\kappa\frac{\alpha}{\rho}\,,
\eeq
and
\beq\label{Hkappa}
H_\kappa(r)= r^2+\kappa +\kappa\frac{\alpha\ell}{r}\,.
\eeq
Besides the sign factor $\kappa$ (with the same meaning as before), these solutions contain two dimensionless parameters, $\alpha$ and $\ell\geq 0$, and all lengths are measured in units of the 3D cosmological radius on the brane, which lies on the section $\rho=\infty$. The metric \eqref{statc2} satisfies
\begin{equation}
    R_{ij}=-3\lp 1+\ell^{-2}\rp g_{ij}\,.
\end{equation}
As we vary $\ell$, the 4D cosmological constant changes too. The central charge $c$ of the dual CFT can nevertheless be kept fixed as we vary $\ell$ by adjusting the effective 3D Newton's constant, $G_3=\ell/(2 c)$ \cite{Emparan:2020znc}. Then when $\ell\ll 1$, the gravitational forces on the brane are weak.

This conclusion that the parameter $\ell$ controls the gravitational backreaction is consistent with the fact that, in the limit $\ell\to 0$, the metric \eqref{statc2}  (renormalized by an overall factor $\ell^2$) becomes the same as \eqref{wickschwads} ($\kappa=-1$) and \eqref{hyperbhads}  ($\kappa=1$) for $n=1$.

In the braneworld construction, we introduce a brane at $\rho=\infty$ and keep the bulk region where $0<\rho<\infty$. The geometry induced on this brane when $\kappa=-1$ gives the chronal, quantum-corrected BTZ black hole with time coordinate $\chi$, which we will study below. When $\kappa=+1$ we get the metric for the quantum-corrected Rindler-AdS$_3$ in the non-chronal region.

In the following, we will focus mostly on the chronal solutions $\kappa=-1$ and with $\alpha>0$. The reason is that in the physical situation we are interested in, we start in a chronologically normal region, and then attempt to evolve to a time machine. We will find that the formation of a chronological horizon is thwarted by the quantum backreaction, hence the non-chronal regions with $\kappa=+1$ are physically unreachable. Nevertheless, it is straightforward to extend our study to include them.

The function $F_-(\rho)$ is independent of $\ell$, and the $(\rho,\phi)$ sector of the metric is conformally equivalent to the one in \eqref{lowerbulk}. Therefore, $\rho$ must be bounded below by the largest positive root, $\rho_0$, of $F_-(\rho)$, and the periodicity of $\phi\sim\phi+\Delta$ must be determined by \eqref{Delrho-} with $n=1$.  The coordinate $\rho$  ranges in $\rho_0\leq \rho\leq \infty$, the latter limit being where the brane lies. Then the $(\rho,\phi)$ sections at constant $\chi$ and $r$, have disk topology. Geometrically, they resemble (roughly) hemispherical `caps', with $\arccos(\rho_0/\rho)$ playing the role of a polar angle.

As explained in \cite{Emparan:1999wa,Emparan:1999fd,Emparan:2020znc}, when $\alpha=0$, we recover empty AdS$_4$, but when $\kappa=-1$ and $\alpha> 0$ there is a black hole stuck to the brane that is qualitatively like (half of) the Schwarzschild-AdS$_4$ black hole---the redshift function $H_-(r)$ has the same form---distorted in the polar direction $\rho$.

\subsection{Curvature singularity}\label{sec:curvsing}

For our present purposes, the most important aspect of this construction is the fate of the chronology horizon when backreaction is included. When the chronal region of Misner-AdS$_3$ is represented as BTZ, \eqref{BTZ}, the chronology horizon lies at $r=0$. We noted that this is a quasi-regular surface that admits an analytic extension across it, and which can be traversed along timelike or null trajectories. The bulk dual shows no singular behavior there---the divergence of the stress tensor being a consequence of the choice of boundary conformal factor. Thus, $r=0$ is also a quasi-regular surface  in the zero-backreaction limit $\ell=0$ of the geometry \eqref{statc2}.

However, when $\ell\neq 0$, we see that as $r\to 0$ the $(\rho,\phi)$-caps in \eqref{statc2} shrink to zero. If $\alpha=0$ (so we recover empty AdS$_4$) this is just an origin of coordinates, but when $\alpha> 0$ we find a curvature singularity at $r=0$, which is spacelike since $r$ is a timelike coordinate there. To see this, we compute the Kretschmann scalar (the square of the Riemann tensor) for \eqref{statc2}, which gives
\begin{equation}
    K = 24 \left(1+\ell ^{-2}\right)^2+\frac{12 \alpha ^2}{\ell^4 } \frac{(\ell +r/\rho)^6}{r^6}.
\end{equation}
When $\alpha,\ell>0$ it diverges for $r\to 0$. This singularity is present for all the range of $\rho$, and in particular on the brane section at $\rho\to\infty$. As we will see, it is qualitatively similar to that of the four-dimensional Schwarzschild black hole. The dependence of $K$ on  $\rho$ simply reflects the non-sphericity of the bulk black hole.

Let us now examine more closely what has happened to the geometry near $r=0$, first in the full bulk spacetime, and then on its brane section. 

\paragraph*{Singularity in the 4D bulk.}  In the absence of backreaction, $\ell=0$, to go near $r=0$  we introduce a time coordinate
\begin{equation}
    \eta = -\ln r
\end{equation}
so when $\eta$ is very large the geometry \eqref{statc2} is approximately given by
\begin{equation}
    ds^2\simeq \rho^2\lp -d\eta^2+e^{2\eta}d\chi^2\rp +\frac{d\rho^2}{F_-(\rho)}+F_-(\rho)d\phi^2\,.
\end{equation}
The geometry of the $(\rho,\phi)$ caps is smooth for all finite $\rho>\rho_0$. The $(\eta,\chi)$ metric is a dS$_2$ space, and $\eta\to \infty$, i.e., $r=0$, is not a singularity but the asymptotic future. This of course corresponds to the observation, made in \cite{Emparan:2021xdy} and in Sec.~\ref{sec:holobulks}, that in the non-backreacted case the chronology horizon lies in the future asymptotic boundary of the bulk.

Now we keep $\ell\neq 0$ and go near $r=0$. To deal with the $(\rho,\phi)$ part of the metric, we take `stereographical' coordinates around any point in the disk, where the metric is approximated by a two-dimensional plane with coordinates $x_2$ and $x_3$. We also introduce the proper time $\eta$ as
\begin{equation}\label{tor0}
    \eta=\frac23 \frac{r^{3/2}}{\sqrt{\mu \ell}}\,,
\end{equation}
and adequately rescale $\chi\to x_1$, so the geometry \eqref{statc2} near $r=0$ is approximated by
\begin{equation}\label{kasner}
    ds^2\simeq -d\eta^2 +\eta^{-2/3}dx_1^2+\eta^{4/3}(dx_2^2+dx_3^2)\,.
\end{equation}
When $\eta\to 0$ this is the same Kasner singularity that is present in the interior of the Schwarzschild solution in four dimensions.

\paragraph*{Singularity in the 3D spacetime.}  The description in terms of the three-dimensional quantum-backreacted spacetime presents some differences. This is the geometry of the slice at $\rho\to\infty$. In the absence of backreaction, we get the boundary geometry of the chronal region of Misner-AdS$_3$ (i.e., BTZ), with the chronology horizon at $r=0$. Renaming $(r,\chi,\phi)=(\eta,x_1,x_2)$ for uniformity of the discussion, the geometry becomes
\begin{equation}
    ds^2\simeq -d\eta^2 +dx_1^2 +\eta^2 dx_2^2\,,
\end{equation}
which is again a Kasner metric but a flat one: $\eta=0$ is a null surface. When $\ell\neq 0$, instead, changing coordinates as in \eqref{tor0} we find a constant-$x_3$ section of \eqref{kasner},
\begin{equation}
    ds^2\simeq -d\eta^2 +\eta^{-2/3}dx_1^2+\eta^{4/3}dx_2^2\,,
\end{equation}
which has a three-dimensional curvature singularity at $\eta=0$. This is not a vacuum Kasner solution (which in three dimensions would have constant curvature), since it is sourced by a quantum CFT with a stress tensor that diverges at $\eta=0$. Actually, the effective three-dimensional theory contains, besides the coupling of the CFT to gravity, an infinite number of higher-curvature corrections to the Einstein-AdS theory \cite{Emparan:2020znc}. These terms can become important near the singularity.

\subsection{Quantum Misner-AdS$_3$}

We finish this section by exhibiting the metric, in the $(\tau,\phi,\chi)$ coordinates, of Misner-AdS$_3$ with the quantum backreaction of the  CFT, in both the chronal and non-chronal regions. This is obtained by simply changing 
\begin{equation}\label{btztomisner}
    r^2=\kappa (\tau^2-1)=|1-\tau^2|
\end{equation}
in the section at $\rho=\infty$, to find
\begin{equation}
   ds^2=-\frac{d\tau^2}{\lp 1-\tau^2\rp\lp 1+\frac{\alpha\,\ell}{\tau^2\sqrt{|1-\tau^2|}}\rp}
    +(1-\tau^2)d\phi^2
    +\tau^2\lp 1 +\frac{\alpha\,\ell}{\tau^2\sqrt{|1-\tau^2|}}\rp d\chi^2\,. \label{qtm}
\end{equation}
When $\ell=0$ this reproduces Misner-AdS$_3$ \eqref{misneradsnpatch}, but when $\ell\neq 0$ we can clearly see the effects of the backreaction on the former chronology horizon at $\tau^2=1$. This was a quasi-regular null surface, but now, in the chronal region, it has become a spacelike curvature singularity (timelike in the non-chronal region). There are no geodesics that can cross it.

We can compute the quantum-corrected stress tensor by solving the Einstein's equations on the brane \eqref{3Deffeqn}. To leading order in $\ell$, in the coordinates $(\tau,\phi,\chi)$, we obtain
\begin{align}
   \langle T_a{}^b\rangle 
   &= \frac{\ell}{16 \pi G_3} \frac{\alpha}{(1 - \tau^2)\sqrt{|1-\tau^2|}}\, \text{diag}(1,-2,1)\,.
\end{align}
If we take into account that $G_3=G/(2\ell)$, and renormalize a factor $\ell^2$, this correctly reproduces \eqref{holostress}. The contributions to the stress tensor to higher orders in $\ell$ can be readily obtained as in \cite{Emparan:2020znc}.

\section{Double holography and non-planar effects of the CFT}\label{sec:doubleh}

\begin{figure}[th]
        \centering
         \includegraphics[width=.9\textwidth]{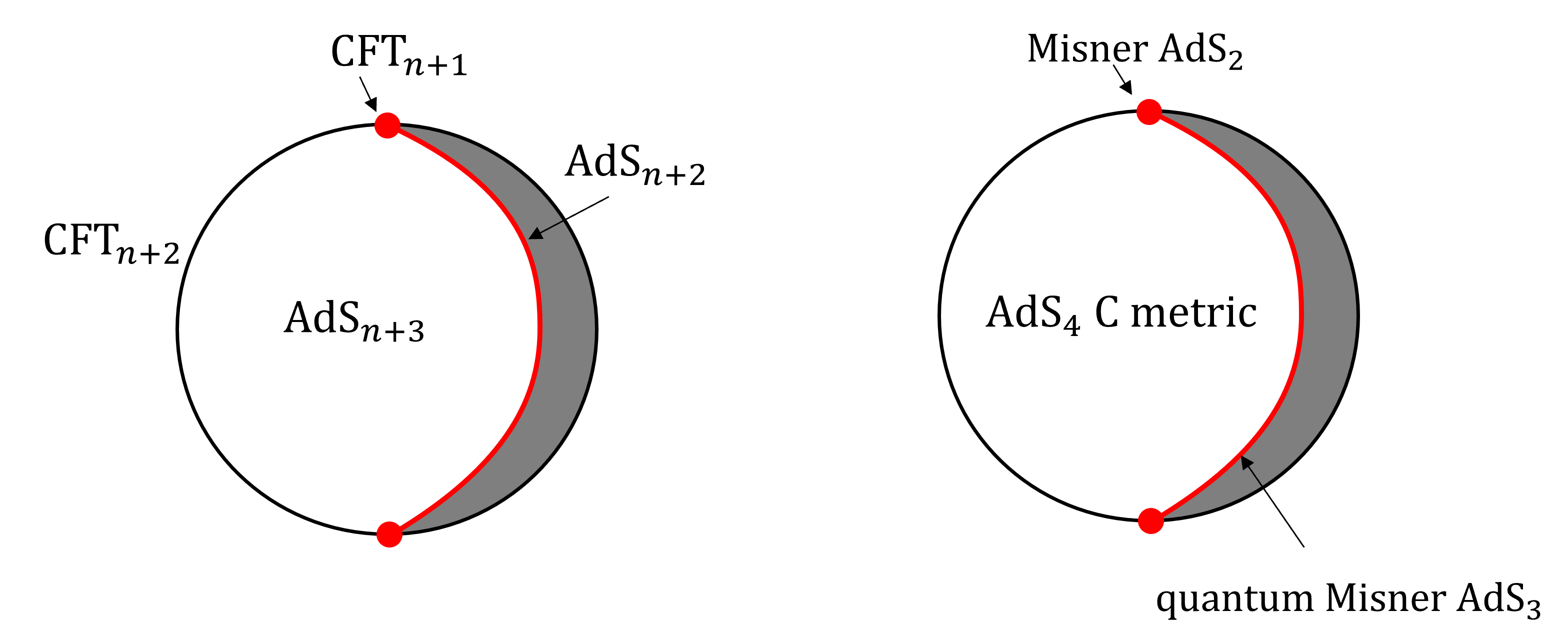}
        \caption{ \small Double holography. The red line is the brane, and the gray region is excluded. Left: general setup ($n\geq 0$). Right: geometries involved in the realization in this article ($n=1$). The classical 4D bulk differs from empty AdS$_4$ and in this way gives rise to quantum bulk corrections to Misner-AdS$_3$ (Sec.~\ref{sec:quMisner}), which in turn are interpreted as $1/c$ corrections to the CFT$_2$ in Misner-AdS$_2$.}
        \label{fig:matrioshka}
\end{figure}

The nesting of an AdS spacetime within a higher AdS spacetime in the manner of \eqref{zeroalpha} allows `doubly holographic' setups through the use of braneworld constructions \cite{Karch:2000ct}. The starting point is an AdS$_{n+3}$ bulk spacetime with a brane that is asymptotically AdS$_{n+2}$. Since the geometry of the brane is dynamical, and described by AdS$_{n+2}$ gravity, it should admit a dual description in terms of a CFT$_{n+1}$ on its (non-dynamical) boundary  (see Fig.~\ref{fig:matrioshka}).
The setup can also be regarded as the holographic realization of a boundary conformal field theory \cite{Takayanagi:2011zk}: the CFT$_{n+2}$ that is dual to the AdS$_{n+3}$ bulk lives on a spacetime that has a boundary, and it is on this boundary that the CFT$_{n+1}$ is localized.

Double holography is of interest to us since, when $n=1$, we can use it to compute the effect of non-planar $1/c$ corrections to a CFT$_2$ in Misner-AdS$_2$. To see how this is possible, recall that the planar solution, i.e., the leading order at large $c$, of the CFT$_2$ is obtained from the classical 3D bulks that we have constructed in Sec.~\ref{sec:holobulks} with $n=0$ (the same as in \cite{Emparan:2021xdy}). Subleading orders correspond to 3D geometries corrected by bulk quantum effects, i.e., with the quantum backreaction of bulk fields.\footnote{Normally, the graviton would be one of these quantum fields, but in 3D there are no gravitons, and even in higher-dimensional setups their contribution is negligible compared to the large number of degrees of freedom of the CFT$_{n+2}$ on the brane.} In the previous section we have seen how to obtain these geometries, and in this way we obtained the quantum-corrected metric \eqref{qtm}.

That is, whereas before we were interested in \eqref{qtm} as the geometry of a quantum-backreacted 3D time machine (Misner-AdS$_3$), we now want to regard it as a bulk geometry that provides the holographic dual of the CFT$_2$ (with $1/c$ corrections) at the Misner-AdS$_2$ boundary. But for this, we have to first rewrite \eqref{qtm} in a form where, indeed, we recover Misner-AdS$_2$ at the asymptotic boundary.

For this purpose, in the following we will consider the case where, at planar order, the CFT$_2$ is in the zero-stress state. This is not only simpler to study than other states, but also more interesting: it is in this instance that a potential violation of chronology protection becomes more pressing. If the stress tensor of the CFT$_2$ does not diverge, what could impede trespassing the chronology horizon? In \cite{Emparan:2021xdy} we claimed that the protection is restored by subleading $1/c$ effects of the CFT$_2$. We will now see how this happens.

The bulk dual of the zero-stress state in Misner-AdS$_2$ is most simply written in the form of \eqref{zeroalpha}, i.e.,
\begin{equation}\label{mads3}
    ds^2=d\sigma^2+\cosh^2\sigma \left( -\frac{d\bar{\tau}^2}{1-\bar{\tau}^2}+\left(1-\bar{\tau}^2\right)d\phi^2\right)\,,
\end{equation}
with periodic $\phi$. This is actually the same spacetime as Misner-AdS$_3$,
\begin{equation}
   ds^2=-\frac{d\tau^2}{1-\tau^2}
    +(1-\tau^2)d\phi^2
    +\tau^2 d\chi^2\,, \label{miads3}
\end{equation}
as follows from the fact that the identification orbits are the same in both cases. But we can also see more explicitly that the change
\begin{equation}\label{mads3tomads3}
    \tau^2=\bar{\tau}^2 \cosh^2\sigma-\sinh^2\sigma\,,\qquad \chi=\arctanh\left(\frac{\bar{\tau}}{\tanh\sigma}\right)\,,
\end{equation}
transforms \eqref{miads3} into \eqref{mads3}.

The metric \eqref{mads3} foliates the geometry in such a way that at the boundaries $\sigma\to\pm\infty$ we recover two copies of Misner-AdS$_2$. Thus, \eqref{mads3} gives the bulk dual to the CFT$_2$ in an entangled state between two copies of Misner-AdS$_2$, at planar order.

As explained in \cite{Emparan:2021xdy} and in Sec.~\ref{sec:holchro} above, the bulk \eqref{mads3} is a traversable wormhole (the two boundaries are disconnected as long as $\bta^2<1$). When traversing it, one crosses the future chronology horizon in the bulk, which in \eqref{mads3} is at $\bta=1$, and in \eqref{miads3} at $\tau=1$ (or $r=0$ in BTZ \eqref{BTZ}). In \cite{Emparan:2021xdy} we found a large class of such bulk wormholes entangling two time machines, which apparently permit violations of chronology. The zero stress state \eqref{mads3} is just the most puzzling of them.

However, when we go beyond the classical limit of the bulk, \eqref{mads3} or \eqref{miads3}, and consider the quantum-backreacted geometry \eqref{qtm}, this passage between boundaries is not possible anymore: now $\tau=1$ is a spacelike curvature singularity and the two boundaries cannot be continuously connected via causal curves through the bulk.\footnote{There is no possible passage through the 4D bulk either: its causal structure is like Schwarzschild-AdS$_4$, which does not have any chronology horizons.}  The same singularity splits into two disconnected components the bulk geometries for the chronal and non-chronal regions of any of the two boundaries, much like in Sec.~\ref{sec:holchro}. In other words, the $1/c$ corrections restore chronology protection to the zero-stress state of the CFT$_2$ in Misner-AdS$_2$. This confirms the claim in \cite{Emparan:2021xdy}.

We can make this point more apparent if we apply to \eqref{qtm} the coordinate transformations \eqref{mads3tomads3}, to show how the classical bulk wormhole \eqref{mads3} becomes non-traversable when the $1/c$ effects of the CFT$_2$ are incorporated. The result is
\begin{align}
ds^2=&\left(
    \frac{(1-\bta^2)\sinh^2\sigma}{f}-\frac{\bta^2 f}{\sinh^4\sigma (1-\bta^2\coth^2\sigma)^2}
    \right)
    \, d\sigma^2
    \nn\\
    &+\bta\sinh 2\sigma \left( \frac{f}{
    \sinh^4\sigma  (1-\bta^2\coth^2\sigma)^2
    }-\frac1{f}
    \right)\,d\tau d\sigma
    \nn\\
    &-\cosh^2\sigma\left(\frac{(1-\bta^2)f}{\sinh^2\sigma (1-\bta^2\coth^2\sigma)^2}-\frac{\bta^2}{f}
    \right)
    \, \frac{d\tau^2}{1-\bta^2}
    \nn\\
    &+\cosh^2\sigma(1-\bta^2)\,d\phi^2\,,\label{mmads3}
\end{align}
where all the dependence on the quantum backreaction parameter $\ell$ is in
\begin{equation}\label{fterm}
    f=(1-\bta^2)\cosh^2\sigma-1-\frac{\alpha\ell}{\cosh\sigma\,\sqrt{|1-\bta^2|}}\,.
\end{equation}
It is easy to verify that at the boundaries $\sigma\to\pm\infty$ we recover Misner-AdS$_2$. When $\ell=0$, that is, without quantum bulk corrections, the metric \eqref{mmads3} reduces (with some work) to \eqref{mads3}. With the corrections for $\alpha\ell\neq 0$, the curvature of \eqref{qtm} diverges at $\tau^2=1$, and so does in \eqref{mmads3} at $\bta^2=1$ too.  At any finite $\sigma$, the chronal bulk is bounded to the future and the past by curvature singularities, which make impossible to cross into the non-chronal region.

One might think that the $1/c$ corrections should generate a non-zero stress tensor diverging at the chronology horizon, associated to the restoration of chronology protection. However, what we have found is subtler. 
The terms in \eqref{mmads3} that are proportional to $\alpha\ell$ decay too quickly towards the boundary to have any effect on the holographic $\langle T_{ab}\rangle$. Therefore, the CFT in this state still has zero stress-energy tensor after including $1/c$ corrections. The modifications of the bulk that close up the chronology horizon, which are given by the last term in \eqref{fterm}, correspond in the CFT to the expectation value of an operator of dimension three that diverges at the chronology horizon. That is, the fact that the field theory is ill-defined on the chronology horizon is not manifest in the stress tensor, but in a different operator of the theory that is associated to its self-interactions.

\section{Outlook}\label{sec:discus}

We have exhibited two mechanisms that quantum physics uses to banish the access to regions with closed timelike curves. The first one only involves the strong self-coupling of the quantum fields, which makes every field excitation interact with a background field state that diverges at the chronology horizon, and then stops it there. Remarkably, the dual bulk geometry is nevertheless everywhere smooth and geodesically complete. This mechanism was uncovered in \cite{Emparan:2021xdy} in two-dimensional Misner spacetimes, and here we have generalized it to arbitrary dimension. We have also shown how the mechanism is strengthened when $1/c$ corrections of the CFT are included, which eliminate the potential violations of chronology that use the entanglement between two time machines. 

The second, more drastic mechanism makes the quantum field backreact on the spacetime geometry, and turns the chronology horizon into an impassable spacelike curvature singularity. We have managed to exhibit this phenomenon with an exact solution (in the planar approximation of the CFT) to the backreaction problem in three dimensions.

How generic are these mechanisms? Misner-type chronology horizons appear in large classes of time machine spacetimes (including, as we have seen, the surface $r=0$ in the static BTZ black hole), but there are notable instances that are not represented by them. Of particular interest are the interiors of spinning black holes, such as the Kerr and rotating BTZ black holes. Their inner horizons are Cauchy horizons beyond which there exist CTCs, but, unlike Misner's horizons, they do not present any obstacle to any geodesics. That is, they are not quasi-regular but fully regular surfaces.

Our holographic constructions make a compelling case that, in a strongly coupled CFT, quantum effects in the leading large $N$ limit suffice to prevent the passage across any quasi-regular chronology horizon. Indeed, we have managed to explicitly present the CFT states (and their bulk duals) that realize this, and account for their singular gravitational backreaction.

The regular Cauchy horizon in rotating BTZ, or in Kerr, seems harder to destroy. Exact solutions for the CFT in the rotating BTZ black hole, to leading order in the large $N$ expansion, still exhibit a smooth Cauchy horizon \cite{Emparan:2020rnp}. Nevertheless, in  \cite{Emparan:2020rnp} we built on the work of \cite{Hollands:2019whz} to argue that higher-order corrections in $1/N$ (or higher loop corrections in non-holographic theories) will render these horizons sufficiently singular to protect the chronological order. This is a satisfying conclusion, but one should keep in mind that in these cases the explicit solution for the backreacted geometries, and the specific nature of the singularity that forms, remain open problems for future work.

\section*{Acknowledgments}
Work supported by ERC Advanced Grant GravBHs-692951, MICINN grants FPA2016-76005-C2-2-P and PID2019-105614GB-C22, and AGAUR grant SGR-2017-754. We also acknowledge financial
support from the State Agency for Research of the Spanish Ministry of Science and Innovation
through the “Unit of Excellence Mar\'{i}a de Maeztu 2020-2023” award to the Institute of Cosmos Sciences (CEX2019-000918-M).

\bibliography{bibi}
\bibliographystyle{utcaps}

\end{document}